\documentclass[10pt,journal,final]{IEEEtran}
\IEEEoverridecommandlockouts

\usepackage{amsfonts,amssymb}
\usepackage{ifpdf}
\usepackage{cite}
\usepackage{multirow}
\usepackage[hyphens]{url}
\usepackage{stfloats}
\usepackage{bm}
\usepackage{color,xcolor}
\usepackage{subfigure}
\usepackage{caption}
\ifCLASSINFOpdf
   \usepackage[pdftex]{graphicx}
   \graphicspath{{../pdf/}{../jpeg/}}
   \DeclareGraphicsExtensions{.pdf,.jpeg,.png}
\else
   \usepackage[dvips]{graphicx}

   \graphicspath{{../eps/}}

   \DeclareGraphicsExtensions{.eps}
\fi
\usepackage{tablefootnote}
\usepackage[cmex10]{amsmath}
\usepackage{algorithm}
\usepackage{algorithmic}
\ifCLASSOPTIONcompsoc
\else
  \usepackage[caption=false,font=footnotesize]{subfig}
\fi
\usepackage{makecell}
\begin{document}

\title{CNNs in the Air via Reconfigurable Intelligent Surfaces}
\author{Meng Hua,~\IEEEmembership{Senior Member,~IEEE,}
 Haotian~Wu,~\IEEEmembership{Member,~IEEE}, and 
 Deniz~G\"und\"uz,~\IEEEmembership{Fellow,~IEEE}
 	\thanks{This work was supported by UKRI under the projects AI-R (EP/X030806/1) and INFORMED-AI (EP/Y028732/1), and by the SNS JU project 6G-GOALS under the EU Horizon program (Grant Agreement No. 101139232). (\emph{Corresponding author: Haotian Wu}.)
 }
 \thanks{The authors are with the Department of Electrical and Electronic Engineering, Imperial College London, London SW7 2AZ, U.K. (e-mail: \{m.hua,haotian.wu17,d.gunduz\}@imperial.ac.uk).}
}

\maketitle
\begin{abstract}
This paper introduces AirCNN, a novel paradigm for implementing convolutional neural networks (CNNs) via over-the-air (OTA) analog computation. By leveraging multiple reconfigurable intelligent surfaces (RISs) and transceiver designs, we engineer the ambient wireless propagation environment to emulate the operations of a CNN layer. 
To comprehensively evaluate AirCNN, we consider two types of CNNs, namely classic two-dimensional (2D) convolution (Conv2d) and light-weight convolution, i.e., depthwise separable convolution (ConvSD). For Conv2d realization via OTA computation, we propose and analyze two RIS-aided transmission architectures: multiple-input multiple-output (MIMO) and multiple-input single-output (MISO), balancing transmission overhead and emulation performance. We jointly optimize all parameters, including the transmitter precoder, receiver combiner, and RIS phase shifts, under practical constraints such as transmit power budget and unit-modulus phase shift requirements. We further extend the framework to ConvSD, which requires distinct transmission strategies for depthwise and pointwise convolutions. Simulation results demonstrate that the proposed AirCNN architectures can achieve satisfactory classification performance. Notably, Conv2d MISO consistently outperforms Conv2d MIMO across various settings, while for ConvSD, MISO is superior only under poor channel conditions. Moreover, employing multiple RISs significantly enhances performance compared to a single RIS, especially in line-of-sight (LoS)-dominated wireless environments.	
	
\end{abstract}

\section{Introduction}
Reconfigurable intelligent surfaces (RISs), also known as intelligent reflecting surfaces (IRSs), have emerged as a key enabling technology for a wide range of applications in 6G wireless communication systems  \cite{10555049}.  Generally speaking, an RIS is a planar metasurface composed of numerous low-cost passive reflecting elements, such as  varactor diodes. A key feature of RIS technology is its ability to dynamically adjust the amplitudes and/or phase shifts of incident signals in real-time, thereby reflecting them in a controlled manner. Representative studies \cite{li2024stacked,chen2023RISMEC,hua2021intelligent} have demonstrated that RISs can significantly enhance network throughput and reduce overall network energy consumption by optimizing the phase shifts.

Beyond wireless communication, RISs have also been proposed for  computation \cite{wangzhibin2022federated,jiangtao2019over,arslan2022over}. 
By adjusting the RIS phase shifts, an edge server can minimize model parameter aggregation errors. Furthermore, the massive array of reflecting elements within an RIS can be regarded as trainable neurons in  a neural network, thereby enabling tight computation–communication co-design. As a result, RIS technology holds great promise for realizing wireless physical neural networks (WPNNs)  with significantly reduced latency and faster inference \cite{hua2026wireless}. 
Despite this potential, research on RIS-assisted WPNNs remains limited  \cite{liu2022programmable,  stylianopoulos2025over,yangyuzhi2024realizing,Garcia2023irNN,zhang2024radio}.
Early work in \cite{liu2022programmable}  developed multi-layer RIS structures devised for
controlled laboratory environments for performing deep learning tasks. Work in \cite{stylianopoulos2025over} and \cite{yangyuzhi2024realizing}  proposed  RISs that
enable edge inference by modeling the RIS-programmable wireless channel as hidden over-the-air (OTA) artificial neural network layers.   In \cite{Garcia2023irNN} and \cite{zhang2024radio}, the authors utilized RISs to program channel impulse responses,  achieving one-dimensional (1D) and two-dimensional (2D) convolutional neural networks (CNNs), respectively. Although these studies provide important insights into OTA neural computation, extending such designs to CNNs and multi-antenna systems is nontrivial and has been largely unexplored. Particularly,  mapping such multi-channel convolutional operations onto physical-layer transformations poses unique challenges, including how to jointly exploit spatial, frequency, and RIS-assisted degrees of freedom (DoFs) to emulate convolution kernels.

In this paper, we  propose AirCNN, a novel WPNN framework that emulates 2D CNNs through joint optimization of multi-RIS phase shifts, transmitter precoders, and receiver combiners. 
Existing methods in \cite{liu2022programmable,   stylianopoulos2025over,yangyuzhi2024realizing,Garcia2023irNN,zhang2024radio} cannot be directly extended to realize 2D CNNs. Emulating 2D CNNs poses additional challenges, including incompatibility with traditional data architectures, transceiver design constraints, and transmission protocol limitations. To address these challenges, we investigate two types of 2D CNNs: the classic 2D convolution (Conv2d) and depthwise separable convolution (ConvSD). We propose corresponding transmission architectures and protocols for both RIS-assisted multiple-input single-output (MISO) and multiple-input multiple-output (MIMO) systems. A comprehensive comparison between these two RIS-aided systems for physical CNN realization is conducted, focusing on both performance and communication  overhead. Simulation results demonstrate that the proposed methods achieve satisfactory classification accuracy. Moreover, it is shown that multi-RIS setups significantly outperform single-RIS configurations, particularly in line-of-sight (LoS)-dominated wireless channels.

\section{System model}

\begin{figure}[!t]
	\centerline{\includegraphics[width=3in]{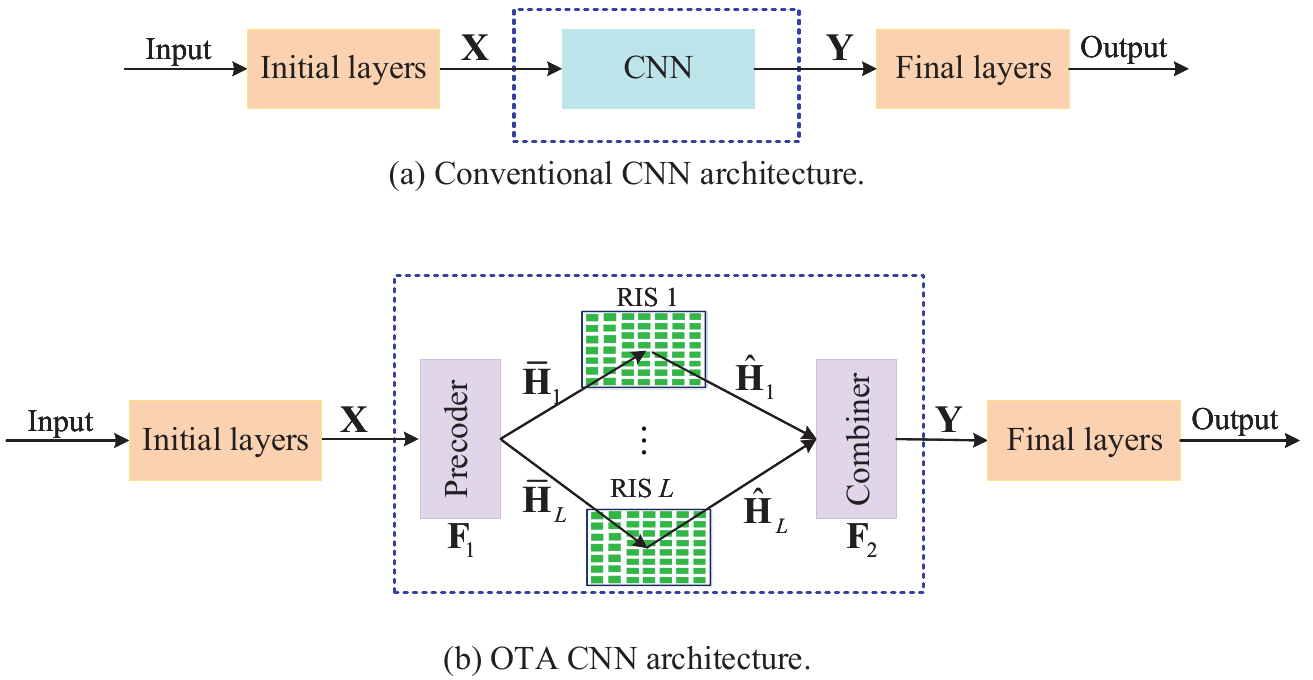}}
	\caption{Illustration of conventional and OTA CNN architectures.}  \label{SystemModel}
	  \vspace{-0.5cm}
\end{figure}

Fig.~\ref{SystemModel}(a) depicts a conventional CNN architecture, which comprises three modules: initial layers, a middle layer, and final layers. The initial and final layers may contain a large number of neural network layers depending on the specific application, while the middle layer corresponds to the CNN. This paper aims to emulate the CNN layer using wireless hardware to achieve a similar function as shown in Fig.~\ref{SystemModel}(b). In the analog-based CNN architecture, the middle layer consists of  precoders at the transmitter, multiple RISs deployed over the air, and  combiners at the receiver.  We assume that the transmitter and receiver are equipped with $N_{\rm t}$ and $N_{\rm r}$ antennas, respectively. Furthermore,  $L$ RISs are deployed, each comprising $M/L$ reflecting elements, where $M$ is the total number of reflecting elements. Let ${{\mathbf{\bar H}}_i} \in {{\mathbb C}^{\left( {M/L} \right) \times {N_{\text{t}}}}}$ and ${{{\mathbf{\hat H}}}_i} \in {{\mathbb C}^{{N_r} \times \left( {M/L} \right)}}$ denote the complex equivalent baseband channel matrices from the transmitter to RIS $i$ and from RIS $i$  to  the receiver, respectively. The  phase-shift matrix of RIS $i$ is denoted by ${{{\bf{\Theta }}_i}} = {\rm{diag}}\left( {{e^{j{\theta _{i,1}}}},{e^{j{\theta _{i,2}}}}, \ldots ,{e^{j{\theta _{i,M/L}}}}} \right)$, with  $\theta_{i,m}$ denoting the $m$-th  phase shift.  

\begin{figure}[!t]
	\centerline{\includegraphics[width=2.5in]{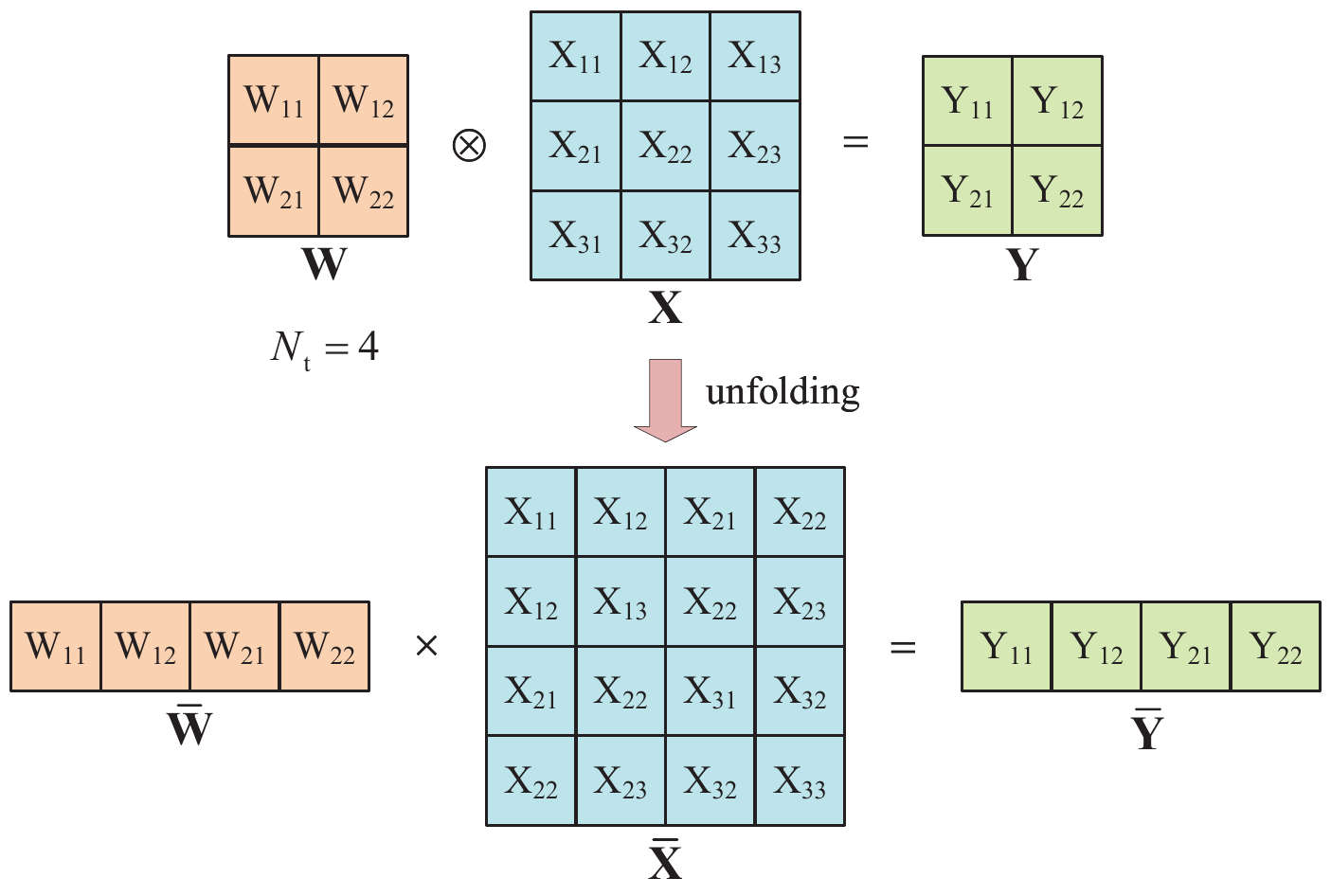}}
	\caption{A toy example of transforming a convolutional operation to a matrix multiplication operation.}  \label{model: toyexample}
	  \vspace{-0.5cm}
\end{figure}
According to the definition of convolution, it is impossible to directly implement CNNs OTA without transformation.
Unlike convolution, matrix multiplication can be inherently realized via OTA transmission.  Therefore, we transform the convolutional operation into a multiplication operation by rearranging the matrices. To clearly illustrate this concept,  Fig.~\ref{model: toyexample} presents a toy example in which  $\bf X$ and $\bf W$ represent a $3\times 3$ input image matrix and a  $2\times 2$ convolutional kernel matrix, respectively. The top part of Fig.~\ref{model: toyexample} shows the standard convolutional operation with a stride of $1$ and no padding, while the bottom part demonstrates the equivalent multiplication matrix operation. Specifically, matrices $\bf W$ and $\bf Y$ are unfolded into vectors $\bf \bar W$ and $\bf \bar Y$, respectively, and rearranged through a piece-wise block vectorization approach. To accommodate practical hardware constraints, the numbers of transmit and receive antennas are set to $4$ and $1$, respectively, matching the unfolded matrix size of $4\times 1$. Thus, after data rearrangement, the convolutional operation can be physically realized. Although Fig.~\ref{model: toyexample} illustrates the case of a single kernel, multiple kernels can be extended similarly.

A key challenge in AirCNN is the joint design of the transmit precoder, receive combiner, and RIS phase-shift matrices to accurately emulate a given digital convolution kernel $\bf \bar W$.  To clarify the formulation,  we consider a case with multiple kernels under single-channel input and single-channel  output. Thus, $\bf \bar W$ is a matrix, where each row represents one kernel. 
This imitation problem can be formulated as\footnote{ The second term in (1a) explicitly characterizes the effect of receiver noise amplified by the post-processing matrix.}
\begin{subequations}  \label{P1}
\begin{align}
    &\mathop {\min }\limits_{{{\bf{F}}_1},{{\bf{F}}_2},{\bf{\Theta }}} \left\| {{{\bf{F}}_2}{{\bf{H}}}{{\bf{F}}_1} - {\bf{\bar W}}} \right\|_F^2 + {{\mathbb E}_{\bf{n}}}\left\{ {{{\left\| {{{\bf{F}}_2}{\bf{n}}} \right\|}^2}} \right\} \label{objectivefunction}\\
   & {\rm{s}}{\rm{.t}}{\rm{. }}~\left\| {{{\bf{F}}_1}} \right\|_F^2 \le {P_{{\rm{max}}}},\\
   &\qquad \left| {{{\left[ {\mathbf{\Theta }} \right]}_{_{i,i}}}} \right| = 1,{\kern 1pt} {\kern 1pt} {\kern 1pt} {\kern 1pt} {\kern 1pt} {\kern 1pt} {\kern 1pt} i = 1, \ldots ,M/L,
\end{align}
\end{subequations}
where ${\bf F}_1$ and ${\bf F}_2$ denote the transmit precoder and receive combiner, respectively,  $P_{\max}$ is the transmit power budget, and $\bf n $ denotes  additive white Gaussian
noise, following ${\mathbf{n}} \sim {\cal CN}\left( {{\bf 0},{\sigma ^2}{\bf I}} \right)$. The end-to-end channel matrix $\bf H$ is modeled as ${\bf{H}} = \sum\limits_{i = 1}^L {{{{\bf{\hat H}}}_i}{{\bf{\Theta }}_i}{{{\bf{\bar H}}}_i}} \in {{\mathbb C}^{{N_r} \times {N_{\text{t}}}}} $.  It is worth noting that problem \eqref{P1} only characterizes the kernel emulation problem for a single input–output channel pair, where multiple convolution kernels are stacked in  $\bf \bar W$. Extensions to practical CNN settings with multiple input and output channels are realized by exploiting time, frequency, and spatial degrees of freedom, as detailed in the subsequent Conv2d and ConvSD MISO/MIMO architectures.

The objective of the proposed AirCNN framework is to perform image classification at the receiver, and all trainable components are optimized in an end-to-end manner.
	Specifically, a standard cross-entropy loss is adopted to train the proposed AirCNN architecture, which can be expressed as
	\begin{align}
	{{\cal L}_{{\text{loss}}}}\left( {{\mathbf{\Omega }},{{\mathbf{F}}_1},{{\mathbf{F}}_2},{\mathbf{\Theta }}} \right) =  - \sum\limits_{i = 1}^C {{p_i}\log \left( {{{\hat p}_i}} \right)} ,
	\end{align}
	where ${\mathbf{\Omega }}$ denotes the digital neural network parameters, $C$ denotes the number of classes, and  ${{p_i}}$ and ${{{\hat p}_i}}$ represent the true label and predicted  probability of the $i$th class, respectively.

\section{Conv2d Realization via OTA Computation}
In this section, we design a physical neural network architecture based on classic Conv2d and propose two realization paradigms: RIS-aided MISO and RIS-aided MIMO systems. Let ${\bf X} \in {\mathbb C}^{B \times C_{\rm in} \times N_{\rm w} \times N_{\rm h}}$  and ${\bf Y} \in {\mathbb C}^{B \times C_{\rm out} \times N_{\rm w} \times N_{\rm h}}$ denote the input and output matrices of the CNN, respectively, as illustrated in  Fig.~\ref{SystemModel}, where $B$ is the batch size, $C_{\rm in}$ and $C_{\rm out}$ denote the numbers of input and output channels, and $N_{\rm w} \times N_{\rm h}$ denotes the input dimensions. The convolutional kernels are assumed to have dimensions of $N_{\rm k}\times N_{\rm k}$. 
\subsection{Conv2d MISO}
For MISO systems, i.e., $N_{\rm r}=1$, we employ time-division multiple access, where one output channel is received per time slot. Specifically,  $C_{\rm in}$ orthogonal frequency division multiplexing (OFDM) carriers are employed at each time, enabling the simultaneous transmission of $C_{\rm in}$ input channels from the transmitter to the receiver. At each time slot $t$, there are $C_{\rm in}$ kernels $\{{\bf \bar w}_{i,t}\}$ to emulate, where
\begin{align}
   {{{\mathbf{\bar w}}}_{i,t}} = {f_{2,i,t}}{\mathbf{h}}_t^H{{\mathbf{F}}_{1,i,t}},i=1,\ldots ,C_{\rm in}, t=1,\ldots ,C_{\rm out}, \label{Conv2dMISO:equation1}
\end{align}
where ${{\mathbf{F}}_{1,i,t}} \in {{\mathbb C}^{{N_{{\text{k}}}^2} \times {N_{{\text{k}}}^2}}}$ and ${f_{2,i,t}} \in {\mathbb C}$ denote the $i$-th OFDM carrier-based precoder at the transmitter and  the amplification coefficient at the receiver, respectively. In addition, ${{\mathbf{h}}_t^H} = \sum\limits_{l = 1}^L {{\mathbf{\hat h}}_l^H{{\mathbf{\Theta }}_{l,t}}{{{\mathbf{\bar H}}}_l}}$, where ${{\bf{\Theta }}_{l,t}}$ represents the $l$th RIS phase shift matrix at time slot $t$. 

At the receiver,  the outputs of 
$C_{\rm in}$ channels are piece-wise summed to generate one output channel at each time slot $t$
\begin{align}
  {{\mathbf{y}}_t} = \sum\limits_{i = 1}^{{C_{{\text{in}}}}} {\left( {{{{\mathbf{\bar w}}}_{i,t}}{{{\mathbf{\bar X}}}} + {f_{2,i,t}}{{\mathbf{n}}_{i,t}}} \right)} , t=1,\ldots ,C_{\rm out}, \label{Conv2d:outputchannel}
\end{align}
where ${\bf n}_{i,t}$ denotes the corresponding  noise term.
After $C_{\rm out}$ transmission time slots, we obtain $C_{\rm out}$  output channels as described in \eqref{Conv2d:outputchannel}. To further enhance transmission efficiency, we dynamically adjust the RIS phase-shift matrix for each time slot, thereby providing more DoFs to emulate the convolution kernels by altering the wireless channels. 
It is important to note that in this system design, each OFDM carrier is associated with a dedicated precoder at each time slot. Consequently, $C_{\rm in}$
different precoders are employed per time slot, while no combiner is needed at the receiver.
\subsection{Conv2d MIMO}
For MIMO systems, we adopt $C_{\rm out}$ receive antennas at the receiver, allowing each antenna to directly capture a distinct output channel. Meanwhile,  $C_{\rm in}$ OFDM carriers are still used for transmitting  $C_{\rm in}$ input channels to each receive antenna. 
Specifically, the relationship can be formulated as
\begin{align}
    {\bf{\bar W}}_i={\bf{F}}_{2,i}{{\bf H}}{{\bf{F}}_{1,i}},i=1,\ldots ,C_{\rm in}, \label{conv2dMIMO:equation1}
\end{align}
where ${{\mathbf{F}}_{1,i}} \in {{\mathbb C}^{{N_{{\text{k}}}^2} \times {N_{{\text{k}}}^2}}}$ and ${{\mathbf{F}}_{2,i}} \in {{\mathbb C}^{{C_{{\text{out}}}} \times {C_{{\text{out}}}}}}$ denote the $i$-th OFDM carrier-based precoder and combiner for convolving with input image ${{{{\mathbf{\bar X}}}}}$, respectively, and ${\bf{H}} = \sum\limits_{i = 1}^L {{{{\bf{\hat H}}}_i}{{\bf{\Theta }}_i}{{{\bf{\bar H}}}_i}} $. Since only a single time slot is required for transmission in this setup, the RIS phase-shift matrices need to be adjusted only once, thereby significantly reducing signaling overhead compared to the MISO scheme.

At each receive antenna,  $C_{\rm in}$ received channels are summed to generate one output channel as 
\begin{align}
   {{\mathbf{Y}}} = \sum\limits_{i = 1}^{{C_{i{\text{n}}}}} {\left( {{{{\mathbf{\bar W}}}_i}{{{\mathbf{\bar X}}}} + {{\mathbf{F}}_{2,i}}{{\mathbf{N}}_i}} \right)},
\end{align} 
where ${\bf N}_i$ denotes the noise vector associated with the 
$i$-th OFDM carrier. 
Unlike the Conv2d MISO scheme, the Conv2d MIMO design requires the use of $C_{\rm in}$ combiners  at the receiver.


\textbf{\textit{Remark:}} In AirCNN, computation is carried out by physical signal propagation rather than sequential digital multiply–accumulate (MAC) operations. Consequently, the dominant complexity does not stem from digital computations, which is also a key motivation for studying WPNNs, but from \textbf{communication-related overhead}, including OTA transmission resources and control signaling.

The communication-related overheads for Conv2d MISO and Conv2d MIMO are summarized  in Table~\ref{Table: Conv2d}. In the table, $T_{\rm s}$ denotes the
	  number of transmission slots, $T_{\rm r}$ the number of RIS adjustments,  $T_{\rm o}$ the number of OFDM carriers, $T_{\rm p}$ the number of precoder adjustments, and $T_{\rm c}$ the number of combiner adjustments.

\begin{table}[!t]
	\centering
	\caption{   Conv2d MISO vs MIMO.}
	\begin{tabular}{|c|c|c|c|c|c|c|c|}
		\hline
            \quad \textbf{Conv2d} & $\bf N_{\rm t}$ & $\bf N_{\rm r}$   &  $\bf T_{\rm s}$ &   $\bf T_{\rm r}$ &$\bf T_{\rm o}$ &$\bf T_{\rm p}$ & $\bf T_{\rm c}$    \\  \hline
		\quad  \textbf{MISO} &$N_{\rm k}^2$  & 1 &$C_{\rm out}$       &$C_{\rm out}$& $C_{\rm in }$            & $C_{\rm out} C_{\rm in}$    & 0   \\ \hline
		\quad  \textbf{MIMO} &$N_{\rm k}^2$  & $C_{\rm out}$  &1       &1 & $C_{\rm in }$            & $C_{\rm in }$    & $C_{\rm in }$  \\ \hline
	\end{tabular}
    \label{Table: Conv2d}
\end{table}

\section{ConvSD Realization via OTA Computation}
In this section, we study the lightweight  ConvSD architecture. ConvSD decomposes the convolution process into two stages: depthwise convolution and  pointwise convolution \cite{chollet2017xception}. In the depthwise convolution, a single convolution filter is applied per input channel,  isolating spatial filtering from inter-channel interactions. In the pointwise convolution, a $1\times1$ convolution is applied to linearly combine the outputs of all depthwise convolutions across channels, thereby creating new feature representations. As a result, the total number of parameters required for ConvSD is $C_{\rm in}\times N_{\rm k}^2+C_{\rm in}\times C_{\rm out}$, which is significantly  fewer than that required for a standard Conv2d operation with $C_{\rm in}\times N_{\rm k}^2\times C_{\rm out}$ parameters.

\subsection{ConvSD MISO}
For the MISO system,  $C_{\rm in}$ OFDM carriers are adopted. Each OFDM carrier is assigned a dedicated precoder (i.e.,  $C_{\rm in}$ different precoders) with each carrier responsible for transmitting one input channel. Since $C_{\rm in}$ OFDM carriers are transmitted simultaneously, only a single transmission slot is needed, and the RIS phase-shift matrix requires adjustment only once.
Mathematically, the operation can be expressed as
\begin{align}
   {{{\mathbf{\bar w}}}_{i}} = {f_{2,i}}{\mathbf{h}}^H{{\mathbf{F}}_{1,i}},i=1,\ldots ,C_{\rm in}, 
\end{align}
which is structurally  similar to \eqref{Conv2dMISO:equation1}.
After receiving the $C_{\rm in}$ distorted input channels corrupted by noise and channel fading,  the receiver applies $C_{\rm out}$ pointwise convolutional filters, each   consisting of  $C_{\rm in}$ kernels of size $1\times1$,  in order to emulate the pointwise convolution step.  Since $N_{\rm r}=1$ in this MISO setting, no combiners are needed at the receiver. The overall operation at the receiver can be formulated as
\begin{align}
   {{\mathbf{y}}_t} = \sum\limits_{i = 1}^{{C_{{\text{in}}}}} {{q_{t,i}}\left( {{{{\mathbf{\bar w}}}_i}{{{\mathbf{\bar X}}}} + {f_{2,i}}{{\mathbf{n}}_i}} \right)} ,t=1,\ldots ,C_{\rm out}, 
\end{align}
where ${{q_{t,i}}}$ denotes the $i$-th kernel coefficient corresponding to the $t$-th output channel. 
\subsection{ConvSD MIMO}
For  MIMO systems, we set $N_{\rm r}=C_{\rm in}$, indicating each receive antenna is responsible for one input channel.
  A combiner of size  $C_{\rm in} \times C_{\rm in}$ is adopted at the receiver, jointly designed with the RIS and the precoder to emulate the depthwise convolution.  Thus, the overall operation can be expressed as
\begin{align}
    {\bf{\bar W}}={\bf{F}}_{2}{{\bf H}}{{\bf{F}}_{1}},
\end{align}
which is similar to \eqref{conv2dMIMO:equation1}, except that only a single precoder and a single combiner are required in this case.
Then, the receiver applies $C_{\rm out}$ filters, each with $C_{\rm in}$ kernels of size $1\times1$, to emulate the pointwise convolution process. We have 
\begin{align}
    {{\mathbf{Y}}_t} = \sum\limits_{i = 1}^{{C_{{\text{in}}}}} {{q_{t,i}}{{\left[ {{\mathbf{\bar W\bar X}} + {{\mathbf{F}}_2}{\mathbf{N}}} \right]}_{i,:}}} ,t=1,\ldots ,C_{\rm out}.
\end{align}
Note that only one precoder and one combiner are adopted in this case.
\begin{table}[!t]
	\centering
	\caption{   ConvSD MISO vs MIMO.}
	\begin{tabular}{|c|c|c|c|c|c|c|c|}
		\hline
		\quad \textbf{ConvSD} & $\bf N_{\rm t}$ & $\bf N_{\rm r}$   &  $\bf T_{\rm s}$ &   $\bf T_{\rm r}$ &$\bf T_{\rm o}$ &$\bf T_{\rm p}$ & $\bf T_{\rm c}$    \\  \hline
		\quad \textbf{ MISO} & $N_{\rm k}^2$  & 1 &1       &1 & $C_{\rm in }$          & $C_{\rm in }$  & 0   \\ \hline
		\quad \textbf{MIMO} &$N_{\rm k}^2$  & $C_{\rm in }$ &1       &1 & 1           & 1    & 1 \\ \hline
	\end{tabular}
	\label{Table: ConvSD}
\end{table}

A communication-related overhead comparison between the parameters of ConvSD MISO and ConvSD MIMO architectures is presented in Table~\ref{Table: ConvSD}.

\section{Numerical results}
In this section, we present numerical results to evaluate the image classification accuracy achieved by the proposed schemes, based on the Fashion MNIST dataset. 
The initial layers consist of one  convolutional (Conv) layer, one real-to-complex (R2C) layer, one batch normalization (BN) layer, and one 
 max pool layer, while the final layers have the similar modules, but with an additional complex-to-real (C2R) layer,
 as the initial layers.
We adopt the Adam optimizer with an initial learning rate of $5\times 10^{-4}$, which is decayed by a factor of 0.8 using a multiplicative learning-rate scheduler. The training batch size is fixed to 64. During both training and testing, a new independent channel realization is drawn for each transmitted image, ensuring evaluation under fast-fading conditions.
In addition, the Rician fading channel with Rician factor $K$ is considered for both MIMO and MISO systems. Unless otherwise specified, we set $C_{\rm in }=32$, $C_{\rm out }=64$, $ N_{\rm wi}=N_{\rm hi}=14$, $ N_{\rm k}=3$, $K= 3~{\rm dB}$, $N_t=9$, $N_r=32$ for ConvSD MIMO,  $N_r=64$ for Conv2d MIMO, $P_{\rm max}=10~{\rm dB}$, and $\sigma^2=1$.


\begin{figure}[!t]
	\centerline{\includegraphics[width=3in]{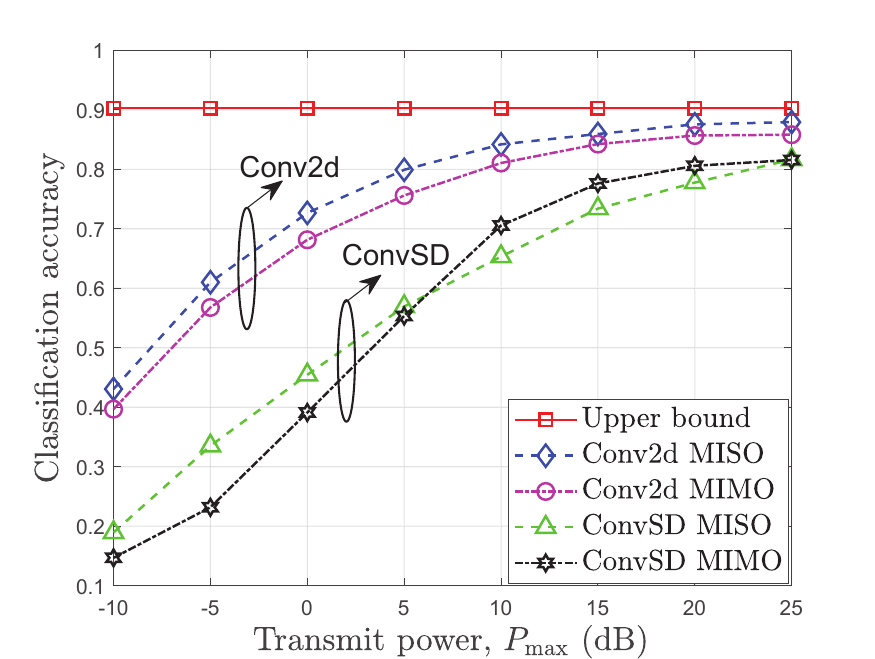}}
	\caption{Transmit power $P_{\rm max}$  versus classification accuracy.}  \label{Simul:PowerVSCA}
	    \vspace{-0.5cm}
\end{figure}

In Fig.~\ref{Simul:PowerVSCA}, we study transmit power $P_{\rm max}$  versus classification accuracy for $L=1$, $K=3~{\rm dB}$, and $M=100$. The ``Upper bound" scheme denotes the digital-domain Conv2d without OTA computation. It can be observed that the classification accuracy of all the schemes except the ``Upper bound" increases with $P_{\rm max}$. This is expected, as higher $P_{\rm max}$  reduces the detrimental impact of noise at the receiver, thereby gradually approaching the ``Upper bound" performance.  Additionally, it is observed that the Conv2d-based schemes consistently outperform the ConvSD-based schemes. This is because the ConvSD layer is a simplified version of the Conv2d layer with weaker feature extraction capabilities. Furthermore, Conv2d MISO consistently outperforms Conv2d MIMO. 
This is because the numbers of adjustments for the precoder and RIS  are $C_{\rm in}C_{\rm out}$ and $C_{\rm out}$, respectively,  whereas they are
$C_{\rm in}$ and $1$ for the Conv2d MIMO scheme, meaning that the DoFs available for emulation in the former scheme are much larger. It should be noted that this observation does not hold for the ConvSD scheme. The ConvSD MISO scheme outperforms the ConvSD MIMO scheme only when the transmit power is low, i.e., below $5~{\rm dB}$,  but performs worse when the transmit power exceeds  $5~{\rm dB}$. This is because the ConvSD MISO scheme adjusts the precoder $C_{\rm in}$ times without adjusting the combiner, whereas the ConvSD MIMO scheme adjusts both the precoder and the combiner once, thus striking different balances. 

\begin{figure}[!t]
	\centerline{\includegraphics[width=3in]{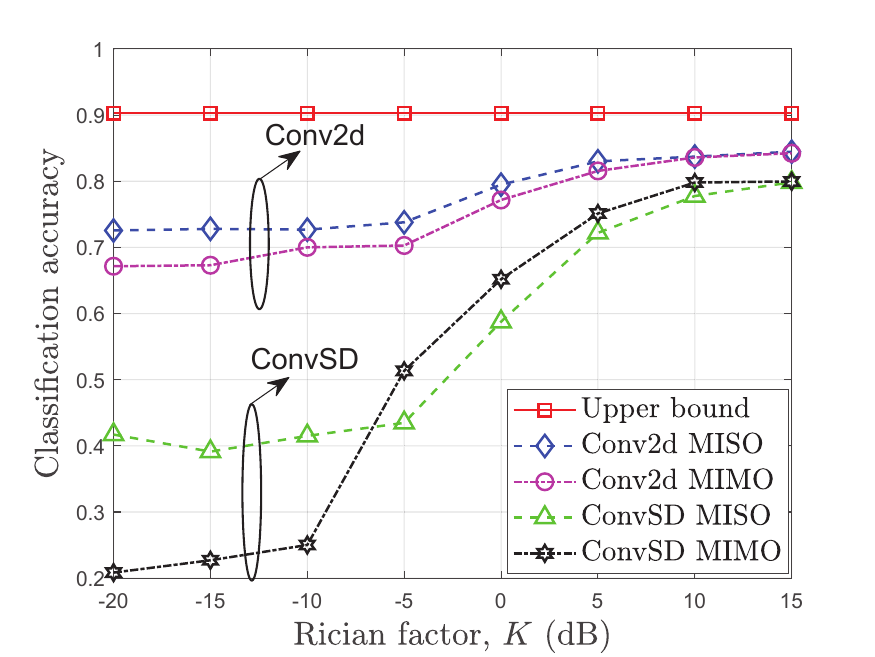}}
	\caption{Rician factor $K$  versus classification accuracy.}  \label{Simul:RicianVSCA}
	    \vspace{-0.5cm}
\end{figure}
\begin{figure}[!t]
	\centerline{\includegraphics[width=3in]{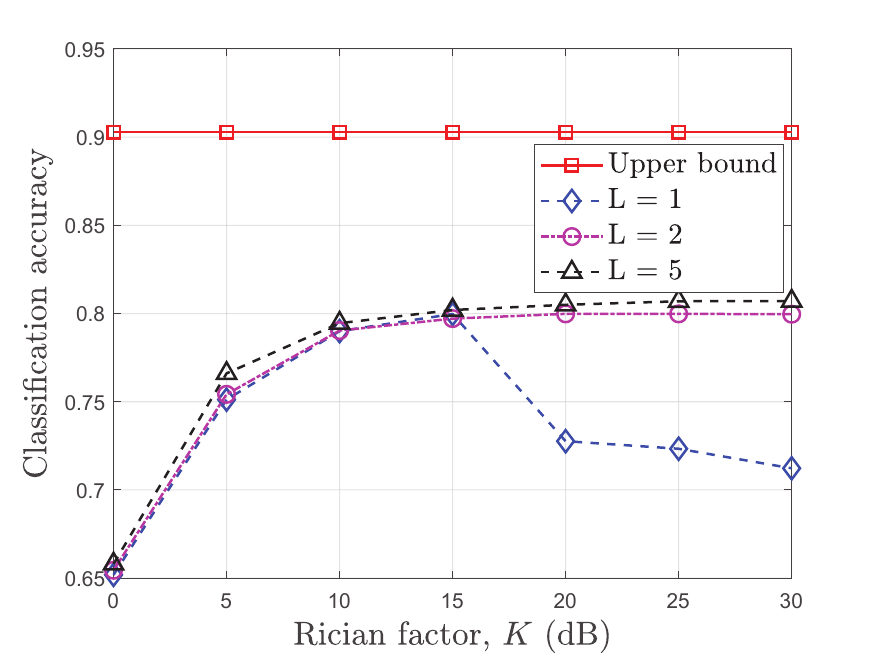}}
	\caption{Rician factor $K$  versus classification accuracy.}  \label{Simul: different No. of RISs}
    \vspace{-0.5cm}
\end{figure}

In Fig.~\ref{Simul:RicianVSCA}, we study Rician factor $K$  versus classification accuracy for $L=1$, $P_{\rm max}=10~{\rm dB}$, and $M=50$. 
As $K$ increases, the channel gain improves, resulting in less signal distortion and enhanced classification accuracy. Furthermore,  when $K$ is below $-10~{\rm dB}$, the ConvSD MISO scheme outperforms the ConvSD MIMO scheme, however, as  $K$ increases, ConvSD MIMO scheme eventually surpasses the ConvSD MISO scheme. 
This behavior is consistent with the explanation provided for  Fig.~\ref{Simul:PowerVSCA}. It is also noteworthy that further increase in $K$ does not always lead to continuous improvements in classification accuracy.
Beyond a certain threshold, increasing $K$ even degrades the performance, which is discussed in Fig.~\ref{Simul: different No. of RISs}.

\begin{figure}[!t]
	\centerline{\includegraphics[width=3in]{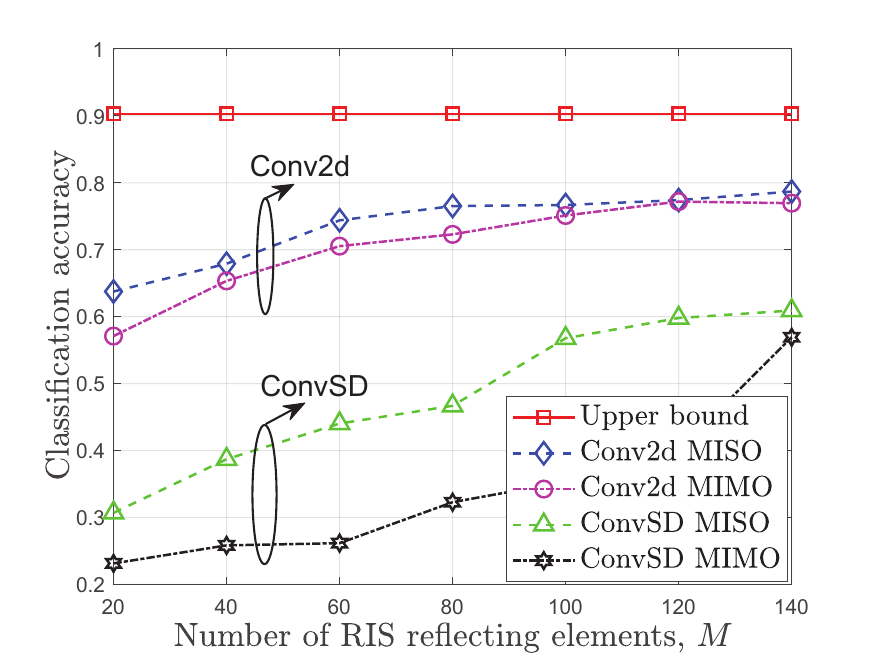}}
	\caption{Number of reflecting elements $M$  versus classification accuracy.}  \label{Simul:RISVSCA}
	    \vspace{-0.5cm}
\end{figure}
In Fig.~\ref{Simul:RISVSCA}, we investigate the classification accuracy  versus the number of  RIS reflecting elements $M$   for $L=1$, $P_{\rm max}=10~{\rm dB}$, and $K=-10~{\rm dB}$.  
It is observed that the classification accuracy of both Conv2D and ConvSD schemes significantly improve with increasing 
$M$. This trend can be attributed to two main reasons. First, a larger number of reflecting elements provides more DoFs for modifying the physical neural network. Second, more reflecting elements enhance the beamforming gain, thus mitigating the impact of noise and improving the classification accuracy.

In Fig.~\ref{Simul: different No. of RISs}, we study the  classification accuracy of the ConvSD MIMO scheme versus $K$ under $P_{\rm max}=10~{\rm dB}$ and $M=50$.  It can be observed that for $L=1$, the classification accuracy  initially increases  but eventually  decreases as $K$  continues to grow. This can be explained as follows. 
When $K$ is below $15~{\rm dB}$, the channel is dominated by non-line-of-sight (NLoS), resulting in a high channel rank and abundant DoFs for neural network modification. Thus,  increasing  $K$ improves classification accuracy by enhancing the channel gain.  However,  when $K \ge 15~{\rm dB}$,  the channel becomes dominated by LoS components. Although the channel gain remains high, the channel rank tends to decrease toward one, limiting the available DoFs, and thereby, degrading the performance.
Moreover, we observe that for larger values of $L$, the system performance remains robust even as $K$ increases. This is because a larger $L$ yields a higher effective channel rank, enhancing the DoFs available for end-to-end training. 

\section{Conclusion}
In this paper, we studied RIS-aided MISO and MIMO systems for engineering the ambient wireless channel to implement CNNs via OTA computation. By jointly training the precoder, combiner, and RIS phase-shift matrices, the digital convolutional operation can be  effectively emulated using physical neural networks.
We investigated two types of CNNs, namely Conv2d and ConvSD, and proposed two transceiver architectures: RIS-aided MISO and RIS-aided MIMO. A comprehensive comparison between these two architectures for physical CNN realization was conducted, highlighting the trade-offs between performance gains and communication overhead.
Simulation results demonstrated that the proposed architectures can achieve satisfactory classification accuracy while enabling OTA computation. Furthermore, it was shown that multi-RIS deployments significantly outperform single-RIS when LoS propagation dominates the wireless channel.

\bibliographystyle{IEEEtran}
\bibliography{AirCNN_Journal}
\end{document}